%% file: main.tex
\documentclass[runningheads]{llncs}
\usepackage{graphicx}
\usepackage[colorlinks,
            linkcolor=red,
            anchorcolor=green,
            citecolor=blue
            ]{hyperref}
\usepackage{mathtools}
\usepackage{adjustbox}
\usepackage{amsmath,array}
\usepackage{multirow}
\usepackage{url}
\usepackage{latexsym}
\usepackage{array}
\usepackage{todonotes}
\usepackage{algorithm}
\usepackage{algorithmic}
\usepackage{makecell}
\usepackage{booktabs}
\usepackage{listings}
\usepackage{amssymb}
\usepackage{bm}
\usepackage{xspace}
\usepackage{cleveref}
\crefformat{section}{\S#2#1#3}
\crefformat{subsection}{\S#2#1#3}
\usepackage{enumitem}
\usepackage{caption}
\usepackage{subcaption}
\usepackage{microtype}
\usepackage{soul}
\usepackage{amsfonts}
\newcommand{\subscript}[2]{$#1 _ #2$}
\usepackage{xcolor,colortbl}
\input{math_commands}

\newcommand\scl{\textsc{SCL}\xspace}
\newcommand\cl{\textsc{CL}\xspace}
\newcommand\modelsclcl{\textsc{Model~+~\scl}\xspace}
\newcommand\modelscl{\textsc{Model~+~\scl+~\cl}\xspace}
\newcommand\tmall{\textsc{Tmall}\xspace}
\newcommand\nowplaying{\textsc{Nowplaying}\xspace}
\newcommand\diginetica{\textsc{Diginetica}\xspace}
\newcommand\dhcn{\textsc{$S^{2}$-$\text{DHCN}$}\xspace}
\newcommand\cotrec{\textsc{COTREC}\xspace}
\newcommand\gcegnn{\textsc{GCE-GNN}\xspace}

\definecolor{cid}{HTML}{dae8f5}
\definecolor{ccon}{HTML}{fee9d4}
\definecolor{gred}{HTML}{cc0200}
\definecolor{ggreen}{HTML}{4C9F26}
\newcommand\blue{\cellcolor{cid}}

\newcommand{\eg}[0]{\emph{e.g., }}

\begin{document}
\title{Self Contrastive Learning for Session-based Recommendation}
%
%

\author{
        Zhengxiang Shi, Xi Wang, Aldo Lipani
}
\institute{University College London, London, United Kingdom \\
        \texttt{\{zhengxiang.shi.19,xi-wang,aldo.lipani\}@ucl.ac.uk} \\
}
\authorrunning{Shi et al.}




%
%
\maketitle              
\begin{abstract}
\input{paper/1_abstract.tex}
\end{abstract}

\input{paper/2_introduction.tex}
\input{paper/7_related_work}
\input{paper/3_background}
\input{paper/4_method.tex}
\input{paper/5_experiment_rq123}

\input{paper/6_experiment_rq4}
\input{paper/8_conclusion.tex}

%
%
%
\clearpage
\bibliographystyle{splncs04}
\bibliography{output}

\end{document}

%% file: math_commands.tex
\usepackage{amsmath,amsfonts,bm}

\def\eqref#1{equation~\ref{#1}}

\def\1{\bm{1}}

\def\va{{\bm{a}}}

\def\vs{{\bm{s}}}

\def\vx{{\bm{x}}}

\DeclareMathAlphabet{\mathsfit}{\encodingdefault}{\sfdefault}{m}{sl}
\SetMathAlphabet{\mathsfit}{bold}{\encodingdefault}{\sfdefault}{bx}{n}



%% file: paper/1_abstract.tex
Session-based recommendation, which aims to predict the next item of users' interest as per an existing sequence interaction of items, has attracted growing applications of Contrastive Learning (CL) with improved user and item representations.
However, these contrastive objectives: 
(1) serve a similar role as the cross-entropy loss while ignoring the item representation space optimisation; and 
(2) commonly require complicated modelling, including complex positive/negative sample constructions and extra data augmentation.
In this work, we introduce Self-Contrastive Learning (\scl), which simplifies the application of CL and enhances the performance of state-of-the-art CL-based recommendation techniques. 
Specifically, \scl is formulated as an objective function that directly promotes a uniform distribution among item representations and efficiently replaces all the existing contrastive objective components of state-of-the-art models.  
Unlike previous works, \scl eliminates the need for any positive/negative sample construction or data augmentation, leading to enhanced interpretability of the item representation space and facilitating its extensibility to existing recommender systems.
Through experiments on three benchmarks, we demonstrate that \scl consistently improves the performance of state-of-the-art models with statistical significance.
Notably, our experiments show that \scl improves the performance of two best-performing models by 8.2\% and 9.5\% in P@10 (Precision) and 9.9\% and 11.2\% in MRR@10 (Mean Reciprocal Rank) on average across different benchmarks.
Additionally, our analysis elucidates the improvement in terms of alignment and uniformity of representations, as well as the effectiveness of \scl with a low computational cost. Code is available at \url{https://github.com/ZhengxiangShi/SelfContrastiveLearningRecSys}.

%% file: paper/2_introduction.tex
\section{Introduction}
\label{sec:introduction}

\begin{figure}[!t]
  \centering
  \includegraphics[width=\columnwidth]{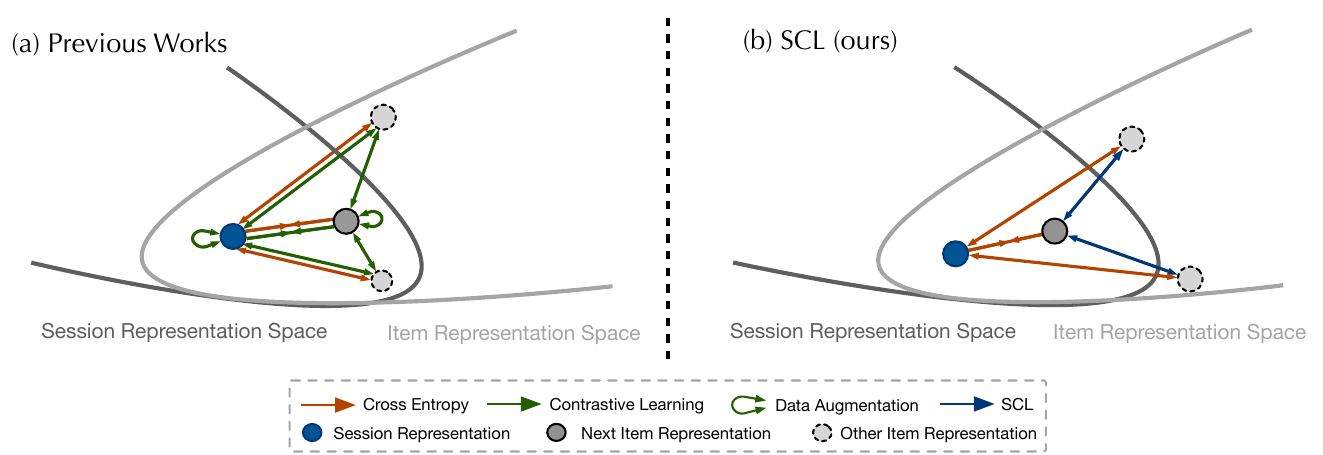}
  \vspace{-1.1em}
  \caption{
  The illustration of the framework of \scl.
  In previous works, contrastive learning (CL) objectives (depicted in green) typically involve complicated modelling, leading to a relatively lesser emphasis on optimising the item representation space.
  In contrast, \scl (depicted in blue) specifically addresses this issue and provides a better complement to the role of cross-entropy loss.
  }
  \label{fig:idea}
\end{figure}

Session-based recommendation \cite{li2017neural,liu2018stamp,xia_self-supervised_2021,xia_self-supervised_2021-1} is a crucial aspect of modern recommender systems for various platforms such as e-commerce websites \cite{jannach2017session,hendriksen2020analyzing}, music streaming services \cite{brost2019music}, and social media \cite{song2019session},
with the goal of predicting a user's next interest by focusing on their current intent.
%
Recently, Contrastive Learning (CL) \cite{oord2018representation} has been applied in session-based recommendation tasks to enhance recommendation accuracy via improved representation quality, with the goal of aligning the session representation with the next item's representation, while also distinguishing it from other item representations. However, two key limitations exist within these methods.

Firstly, \textbf{the importance of optimising the item representation space \textit{itself} by ensuring that the representations are uniformly distributed is not receiving adequate attention}.
The CL objectives in previous research \cite{10.1145/3511808.3557081,duorec2022,10.1145/3474085.3475665,xia_self-supervised_2021,xia_self-supervised_2021-1,xie2022multi,10.1145/3477495.3531937} serve a similar role as the cross-entropy loss while the optimisation of item representation space is not adequately addressed.
As shown in Figure \ref{fig:idea}(a), both cross-entropy loss and CL loss have the capacity to align the session representation with the representation of the next item and differentiate it from other item representations (\cref{sec:movtivation}).
This has led to the marginalization of the role played by the uniformity of item representations.
While some studies \cite{duorec2022,10.1145/3477495.3531937,10.1145/3511808.3557081,10.1145/3474085.3475665,xie2022multi} have touched upon improving the uniformity of representations, these efforts generally contribute only a small fraction to the overall loss function.

\looseness -1 Secondly, \textbf{current CL-based approaches often utilise complex techniques, including the sophisticated creation of positive and negative pairs and extra data augmentations, leading to limited adaptability across models}. Specifically, two state-of-the-art (SOTA) session-based recommendation models, \dhcn \cite{xia_self-supervised_2021-1} and \cotrec \cite{xia_self-supervised_2021} are the typical examples of complex CL-based applications. 
\dhcn encompasses two encoder networks that generate varied session representations (positive) and compare them to corrupted session representations (negative) for noisy data-augmented CL.
Similarly, \cotrec requires two item representations to interact with the corresponding session representation in the CL objective, obtained through model-specific data augmentation techniques. These methods are heavily dependent on the model architecture and may not be compatible with various other models.
Moreover, while recent studies have highlighted the importance of uniformity in user/item representations for recommendation tasks, this has simultaneously triggered a rise in the use of extra data augmentation methods, such as applying noise perturbation \cite{10.1145/3477495.3531937} or dropout \cite{10.1145/3591469,xie2022multi} to augment representations, as shown in Figure \ref{fig:idea}(a).


\looseness -1 In this work, we argue that the importance of the uniformity of item representations has been considerably undervalued and that intricate CL objectives could be streamlined.
We propose a novel approach, \textit{Self Contrastive Learning} (\scl), which directly enforces the representation of each item distinct from those of all other items through a new loss objective and thus promotes a uniform distribution within the item representation space.
\scl can be easily integrated into SOTA models to effectively replace other CL objectives, eliminating the need for creating complex positive/negative samples or engaging in any form of data augmentation.
Different from previous approaches in recommendation systems that utilise the CL \cite{xie2020contrastive,10.1145/3474085.3475665,10.1145/3511808.3557317,duorec2022,10.1145/3477495.3531937}, \scl represents the first attempt to simply enforce uniformity of item representation without resorting to other CL objectives. 
Through our research, we aim to address the following research questions:
\begin{enumerate}[label=\subscript{\textbf{RQ}}{{\arabic*}}]
    \item \label{Q1} To what extent does \scl improve the session-based recommendation tasks? (\cref{sec:main_results})
    \item \label{Q2} How does \scl improve the model performance in terms of the alignment and uniformity of representations? (\cref{sec:alignment_uniformity})
    \item \label{Q3} Are those sophisticated CL objectives still necessary in the presence of \scl? (\cref{sec:really_matter})
    \item \label{Q4} Can \scl maintain SOTA performance with a low computational cost? (\cref{sec:computation_cost})
\end{enumerate}

To address \ref{Q1}, we experiment on three datasets, \tmall, \nowplaying, and \diginetica (\cref{sec:main_results}). Our results show that \scl consistently improves the performance of SOTA models across various evaluation metrics and datasets. 
In particular, our experiments on \tmall show that \scl improves the performance of \dhcn from 28.65\% to 35.14\% in P@$10$ and from 15.94\% to 20.39\% in MRR@$10$, and it also boosts the performance of \cotrec from 30.44\% to 35.03\% in P@$10$ and from 17.28\% to 20.46\% in MRR@$10$, outperforming all existing approaches by large margins.

To understand how the model is improved (\ref{Q2}), we investigate the transformations of the session and item representation in terms of alignment and uniformity (\cref{sec:alignment_uniformity}). 
Our study reveals that \scl learns item representations with a lower uniformity loss, leading to significant improvements in performance. Our findings suggest that SOTA approaches may have placed excessive emphasis on the alignment of session and item representations.

To answer \ref{Q3}, we carry out an ablation study to evaluate the necessity of sophisticated CL objectives employed in prior works (\cref{sec:really_matter}). Our experiment reveals that \scl is capable of attaining the comparable model performance on its own, suggesting the advance of \scl and the redundant use of existing heavy and sophisticated CL objectives.


To understand the computational efficiency (\ref{Q4}), we further study the impact of selecting the $k$-nearest item representations in \scl on the model performance (\cref{sec:computation_cost}). 
Our results show that \scl generally benefits from contrasting to more item representations. However, it can still achieve SOTA performance even just using a value of $k$ equal to 2, indicating that \scl can be implemented with a low computational cost.

%% file: paper/7_related_work.tex
\section{Related Work}
\paragraph{\textbf{Session-based Recommendation.}}
The session-based recommendation aims to predict the next item by utilizing user behaviours within a short time period \cite{hidasi2015session,liu2018stamp,wu2019session,wang2020beyond}.
Early studies on session-based recommendation focused on utilising temporal information from session data through the use of Markov chain models \cite{shani2005mdp,li2017neural,rendle2010factorizing,zimdars2013using,fu2023,yin2016spatio,xu2019graph}. 
Then neural networks are widely applied for session-based recommendation \cite{10.1145/3539813.3545126,hidasi2015session,liu2018stamp,wu2019session,wang2020beyond}.
Recurrent neural networks \cite{hochreiter1997long} have been applied to session-based recommendation models to capture the sequential order between items \cite{zhang2014sequential}. 
GRU4Rec \cite{hidasi2015session} was the first model to use gated recurrent units (GRUs) \cite{chung2014empirical} to model the sequential relations of item interactions. 
NARM \cite{li2017neural} extended GRU4Rec by incorporating the attention mechanism \cite{bahdanau2014neural,Shi2022learning,Shi2022attention,shi2023and}
%
STAMP \cite{liu2018stamp} also replaced the recurrent structure with attention layers to capture a user's general and current interests.
%
%
%
%
%
FGNN \cite{qiu2019rethinking} rethinks the sequence order of items to exploit users' intrinsic intents using GNNs. 
GCE-GNN \cite{wang2020global} aggregates item information from both the item-level and session-level through graph convolution and self-attention mechanism.
\dhcn \cite{xia_self-supervised_2021-1} utilizes hyper-graph convolutional networks to capture high-order item relations within individual sessions.
\cotrec \cite{xia_self-supervised_2021} integrated self-supervised learning into the graph training through sophisticated positive and negative constructions.

\paragraph{\textbf{Contrastive Learning.}}
CL has achieved great success in various research domains, such as computer vision~\cite{he2020momentum} and natural language processing~\cite{gao2021simcse,10.1145/3539813.3545126,shi2022stepgame,shi2023rethinking,shi2023don}, 
with the goal of obtaining high-quality representations by pulling positive or similar instances closer in the representation space while simultaneously separating dissimilar, or negative instances.
Recently, CL has recently been applied to sequential recommendation tasks, with several studies exploring its potential benefits in this area \cite{yao2020self,yu2021self,yu2021socially}. 
Bert4Rec \cite{sun2019bert4rec} adapts the cloze objective from language modelling to a sequential recommendation by predicting random masked items in the sequence with surrounding contexts. 
$S^{3}$-Rec \cite{zhou2020s} utilizes intrinsic data correlations among attributes, items, subsequences, and sequences to generate contrastive signals and enhance data representations through pre-training.
%
%
\cite{ma2020disentangled} proposed a seq2seq training strategy based on latent self-supervision and disentanglement of user intention behind behaviour sequences. 
CoSeRec \cite{liu2021contrastive} uses GNNs to capture more complex patterns than sequential patterns via CL objectives.
CL4SRec~\cite{xie2022contrastive} combines recommendation loss with CL loss of self-supervised tasks.
DuoRec~\cite{duorec2022} retrieves the positive view of a given user sequence by identifying another user's sequence that shares the same next item through its proposed supervised CL. Unlike DuoRec, our task does not have access to user information.
CL has also been applied to other recommendation paradigms, such as general recommendation \cite{yao2020self} and social recommendation \cite{yu2021self,yu2021socially}.
Previous work \cite{Bae_Kim_Ko_Lee_Noh_Yun_2023} also proposes a framework as self contrastive learning approach where every sample sharing a ground-truth label with an anchor is treated as a positive pair. In contrast, our approach is different because we define the positive sample exclusively as the sample itself, not considering other samples with the same label.
%

%% file: paper/3_background.tex
\section{Preliminaries}
\paragraph{\textbf{Task Definition.}}
In the session-based recommendation, the full set of item candidates is represented as $I=\{i_1, \dots, i_n\}$, where $n$ is the total number of item candidates. A session $s$, consisting of $m$ items, is represented as a sequence $S = [i^s_1, \dots, i^s_m]$ ordered by timestamps, where $i^s_k \in I$ ($1 \leq k \leq m$) represents the $k$-th item that has been interacted with by a user. The objective is to predict the next item, $i^s_{m+1}$, from a full set of item candidates $I$, based on the corresponding session sequence $S$. 
For a session $s$, the output of the session-based recommendation model is a ranked list of item candidates $R=[r^s_1,\dots,r^s_n]$, where $r^s_*$ is the corresponding predicted ranking or preference score of the $i$-th item. Afterwards, the top-$k$ items ($1 \leq k \leq n$) will be selected as recommendations.

\paragraph{\textbf{Contrastive Learning.}}
Contrastive learning aims to pull the representation of an anchor sample and the representations of its corresponding positive sample pairs closer while simultaneously pushing the representations of the negative sample pairs away \cite{gao2021simcse}. 
%
\textsc{InfoNCE} \cite{oord2018representation}, where NCE stands for Noise Contrastive Estimation, is a type of contrastive loss function commonly used in recommendation systems \cite{xie2020contrastive,xie2022contrastive,duorec2022,xia_self-supervised_2021-1}.
Formally, let $\va$ denote an anchor representation and $\mathcal{X} \triangleq \{\vx_1, \ldots, \vx_{n-1}, \vx_{n}\}$ denote the set of negative representations ($1 \leq k \leq n-1$) and one positive representation ($k=n$) with respect to $\va$, the \textsc{InfoNCE} loss is defined as:
\begin{equation}
    \label{eq:infonce}
    \begin{aligned}
        \mathcal{L_{\textsc{InfoNCE}}} = -\log \frac{f(\va, \vx_n)}{\sum_{j=1}^n f(\va, \vx_j)},
    \end{aligned}
\end{equation}
where $f$ can be approximated by a real-valued scoring function and typically a function of the cosine similarity is used.

In the field of CL, two key properties, known as alignment and uniformity, have been proposed by \cite{10.5555/3524938.3525859} as measures of the quality of representations. 
The uniformity of the embedding distribution is measured as follows:
\begin{align}
    \label{eq:uniform}
    \ell_{\text {uniform}} &\triangleq \log  
    \underset{{\substack{x \sim p_{\text{data}}, \\ x' \sim p_{\text{data}}}}}{\mathbb{E}} e^{-2\|f(x)-f(x')\|^{2}},
\end{align}
where $p_\text{data}$ denotes the data distribution.
$\ell_{\text {uniform}}$ is lower when random samples are farther from each other. Therefore, the examination of item representation uniformity ensures their semantic interpretability for a potential improvement in identifying the items of true interest. 
In contrast, instead of assessing the dispersion of item representations for uniformity, alignment gauges the expected distance between the embeddings of positively paired instances, assuming that representations are normalized, as expressed by the following equation:
\begin{align}
    \label{eq:align}
    \ell_{\text{align}} & \triangleq 
    \underset{{\substack{x \sim p_{\text{data}}, \\ x' \sim p_{\text{pos}}(x)}}}{\mathbb{E}}{\left\|f(x)-f(x')\right\|^{2}},
\end{align}
%
where $p_{\text{pos}}(x)$ denotes the data distribution of samples that are positive to the instance $x$. $\ell_{\text{align}}$ is lower as all positive samples are closer to each other.
%

%% file: paper/4_method.tex
\section{Methodology}
In this section, we first discuss the potential limitations of SOTA session-based recommendation systems (\cref{sec:movtivation}).
To address these issues, we then introduce a novel approach, \textit{Self Contrastive Learning} (\scl), which aims to improve the uniformity in item representations by utilising a novel loss function (\cref{sec:scl}).

\paragraph{\textbf{Motivation.}}
\label{sec:movtivation}
In the field of session-based recommendation, existing works \cite{wang2020global,xia_self-supervised_2021,xia_self-supervised_2021-1}, that utilise CL objectives, generally employ a framework in which the loss function is a combination of cross-entropy ($\mathcal{L}_{ce}$) and CL $\mathcal{L}_{cl}$ losses:
\begin{equation}
    \label{equation:current_framework}
    \mathcal{L} = \mathcal{L}_{ce} + \alpha \mathcal{L}_{cl},
\end{equation}
where $\mathcal{L}_{ce}$ aims to maximize the likelihood of selecting the correct next item, and $\mathcal{L}_{cl}$ aims to improve the learned representations, with the scalar coefficient $\alpha$ controlling the relative importance of these two objectives. 
Typically, the \textsc{InfoNCE} loss, as in Eq.~\ref{eq:infonce}, is used as $\mathcal{L}_{cl}$.
However, these two objectives are similar in nature, as shown in Figure \ref{fig:idea}.
Specifically, let $\vs$ denote a learned session representation and $L = \{(\vx_k, y_k)\}^{n}_{k=1}$ denote a set of $n$ learned item representations and their corresponding ground-truth labels, where $y_k$ is 1 if the $k$-th item is the user’s next click item and 0 otherwise.
The categorical cross-entropy of classifying the next item correctly is computed as follows:
\begin{align}
    \label{equation:ce}
    \mathcal{L}_{ce} &= - 
    \sum \limits_{\vx, y \in L} y \log p(\vx|\vs),
\end{align}
where $p$ measures the probability that the item represented by $\vx$ is drawn from the full set of item candidates conditioned on the session representation $\vs$. The probability measure $p$ is typically normalized using a real-valued scoring function $f$ (e.g., cosine similarity). Thus, we can rewrite the Eq.~\ref{equation:ce} as:
\begin{align}
    \label{equation:motivation}
    \mathcal{L}_{ce} &= - 
    \sum \limits_{\vx, y \in L} y \log \frac{f(\vs,\vx)}{\sum_{j=1}^n f(\vs,\vx_j)} 
    = -\log \frac{f(\bm{s},\vx^+)}{\sum_{j=1}^n f(\vs,\vx_j)}, 
\end{align}
where $\vx^+$ is the user's next clicked item. 
Therefore, $\mathcal{L}_{ce}$ can be considered as an alternative expression of $\mathcal{L}_{cl}$ ($\mathcal{L_{\textsc{InfoNCE}}}$) when they use the same function $f$.

It is important to note that, while the loss functions $\mathcal{L}_{ce}$ and $\mathcal{L}_{cl}$ in Eq.~\ref{equation:current_framework} may have marginal variations, (\eg $\mathcal{L}_{cl}$ may use the extra temperature parameter $\tau$ in the function $f$ in Eq. \ref{equation:motivation} and different positive and negative samples from data augmentation), their directions of optimising the representation spaces are the same: both objectives aim to push the session representation $\vs$ closer to the next item representation $\vx^+$ while pulling it away from other representations $\vx_j$, thus improving session and item representations.
Although using \textsc{InfoNCE} as CL objectives in conjunction with cross-entropy loss may result in a marginal improvement in performance, we argue that it is not the most effective strategy. 
Firstly, this may place an overemphasis on the alignment of the session and item representations, as shown in Figure \ref{fig:idea}(a) where the green lines and red lines both have an impact on alignment.
Secondly, while prior work \cite{xie2022multi} attempted to improve the uniformity of the item representation space with some auxiliary losses, the importance of blue lines (see Figure \ref{fig:idea}(b)) appears to be diluted by other CL objectives.
A more straightforward regularization approach that specifically targets the item representation distributions and effectively complements the cross-entropy loss is necessary to improve the overall recommendation performance.


\paragraph{\textbf{Self Contrastive Learning (\scl).}}
\label{sec:scl}
To address the aforementioned issues, we propose Self Contrastive Learning (\scl), a straightforward solution to improve the uniformity of the item representation space by introducing an additional loss objective, as shown in Figure \ref{fig:idea}(b).
This objective operates by directly penalizing the proximity of item representations based on our assumption that the representation of each item representation should be distant from those of all other items.
Formally, given a set of $n$ learned item representations $\mathcal{X}$, the objective of the \scl loss is calculated as follows:
\begin{equation}
    \label{eq:scl}
    \begin{aligned}
        \mathcal{L_{\scl}} = - \sum_{i=1}^n \log \frac{g(\vx_i, \vx_i)}{\sum_{j=1}^n g(\vx_i, \vx_j)},
    \end{aligned}
\end{equation}
where the function $g(\vx,\vx')$ is computed by $e^{\text{sim}(\vx,\vx')/\tau}$, the exponential of the cosine similarity controlled by a temperature parameter $\tau$. 
Using the cosine similarity, this loss pulls apart all items on the unit hypersphere.
%
Next, we integrate $\mathcal{L_{\scl}}$ into the existing session-based recommendation models.
Given the loss objective $\mathcal{L}_{model}$ from the original model 
(\textbf{all other CL objectives are excluded}), the overall loss function is computed as follows:
\begin{equation} 
    \mathcal{L} = \mathcal{L}_{model} + \beta\mathcal{L}_{\scl},
\end{equation}
where $\beta$ is a hyperparameter that determines the relative importance of the two objectives. Complementary to the $\mathcal{L}_{model}$, which typical uses a $\mathcal{L}_{ce}$ to positively impact both $\ell_{align}$ and $\ell_{uniform}$, $\mathcal{L}_{\scl}$ has a stronger positive effect on $\ell_{uniform}$.

The advantages of \scl can be summarized in three main aspects:
(1) \textbf{Improved representation spaces.}
By incorporating \scl as an additional loss objective, we achieve improved uniformity in the item representation space, leading to better model performance;
(2) \textbf{Simplified modelling process.}
By leveraging the \scl objective, we avoid the need for complex creation of positive/negative sample pairs and data augmentation techniques, such as noise perturbation \cite{10.1145/3477495.3531937} or dropout \cite{xie2022multi}. In \scl, each item representation serves as the sole positive sample, and all other item representations are considered negative samples without further modifications. This greatly simplifies the construction of recommendation systems, making them more efficient and easier to implement;
and (3) \textbf{Seamless integration into existing systems.}
\scl can be seamlessly integrated into existing session-based recommendation systems that utilise session and item representations, without any additional modification to the architecture of the model. This high level of compatibility makes \scl widely applicable and adaptable to various settings and scenarios.
Overall, these advantages make \scl a valuable solution for enhancing recommendation systems, offering improved uniformity, simplified training, and easy integration into existing models.

%% file: paper/5_experiment_rq123.tex
\section{Experiments and Results}
\label{sec:rec_without_user_profile}

In this section, we evaluate our \scl in three benchmarks. We first describe the experimental setup, including the used datasets, baselines, evaluation metrics, and implementation details.
Then we present our experimental results with respect to the four research questions introduced in Section \cref{sec:introduction}.

\subsection{Experimental Setup}
\paragraph{\textbf{Datasets and Baselines.}} 
\textbf{\tmall}
is sourced from the IJCAI-15 competition and includes anonymized shopping logs from users on the Tmall online shopping platform, with train size 351,268, test size 25,898, items Size 40,728 and an average length 6.69.
\textbf{\nowplaying}
describes the music-listening behaviour of users, with train size 825,304, test size 89,824, items Size 60,417 and an average length 7.42; and
\textbf{\diginetica}
, from CIKM Cup 2016, comprises typical transaction data, with train size 719,470, test size 60,858, items Size 43,097 and an average length 5.12.
Our proposed \scl method is compared with the following representative methods:
\textbf{FPMC} \cite{rendle2010factorizing},
\textbf{GRU4REC} \cite{hidasi2015session},
\textbf{NARM} \cite{li2017neural},
\textbf{STAMP} \cite{liu2018stamp},
\textbf{SR-GNN} \cite{wu2019session},
\textbf{\gcegnn} \cite{wang2020global},
\textbf{\dhcn} \cite{xia_self-supervised_2021-1}, and
\textbf{\cotrec} \cite{xia_self-supervised_2021}.

\input{table/baseline.tex}


\paragraph{\textbf{Implementation Details.}}
We evaluate the performance of our proposed \scl method using the metrics of P@$k$ (Precision) and MRR@$k$ (Mean Reciprocal Rank), where the cutoff $k$ is set to 5, 10 or 20.
We conduct experiments with the proposed \scl method using three SOTA models, \gcegnn,
\dhcn,
and \cotrec.
We first reproduce the experimental results of these models by following the settings in their original papers.
Then, we apply the \scl to these three models.
For the hyperparameters used in the \scl, the temperature parameter, denoted by $\tau$, is set to 0.1, and the loss weight parameter, denoted by $\beta$, is varied within a range of 0.1 to 100.
We have omitted the evaluation of \cotrec on the \nowplaying as we were unable to replicate the results.



\subsection{Main results (RQ1)}
\label{sec:main_results}

Table \ref{table:experiment_results_1} presents the performance of all comparison methods, where our \scl is applied to three SOTA models, \gcegnn, \cotrec, and \dhcn. 
Results show that \scl consistently improves the model performance in terms of P@$k$ and MRR@$k$ across three datasets, \tmall, \nowplaying, and \diginetica, achieving the new SOTA performance (highlighted in blue). 
The significance tests further corroborate the effectiveness of \scl.
Particularly remarkable is that \scl achieves a notable improvement compared to the SOTA models on the \tmall dataset.
\cotrec model with the proposed \scl method also shows significant improvement, with an increase of 18.4\% and 17.5\% in terms of MRR@$10$ and MRR@$20$, respectively.
Additionally, \dhcn+~\scl achieves a new SOTA performance on the \tmall dataset.
It records a 27.9\% increase of the MRR@$10$ from 15.94\% to 20.39\% and a 26.4\% increase of MRR@$20$ from 16.35\% to 20.67\%.
Similar improvement in performance can also be observed on the \nowplaying and \diginetica datasets.

\begin{figure*}[!t]
  \centering
  \includegraphics[width=\textwidth]{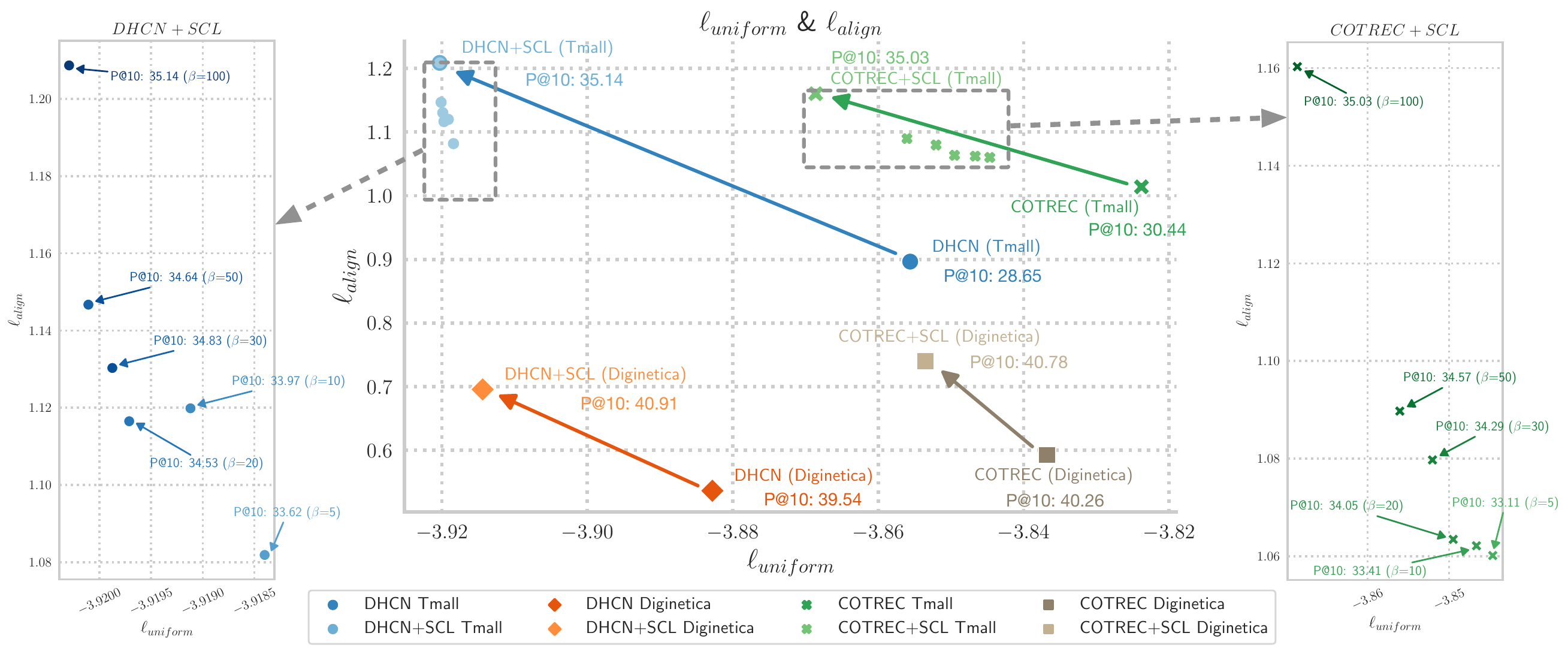}
  \caption{The analysis of alignment loss $\ell_{\text{align}}$ and uniformity loss $\ell_{\text{uniform}}$, where P@$10$ is reported as model performance. 
  While it is generally acknowledged that a decrease in the alignment or uniformity loss leads to improved model performance, an excessive emphasis on alignment and insufficient attention to uniformity can result in sub-optimal performance.}
  \label{fig:alignment_uniformity}
  \vspace{-1em}
\end{figure*}

\subsection{Alignment and uniformity (RQ2)}
\label{sec:alignment_uniformity}
The substantial improvement in performance achieved by the \scl raises the research question of where these improvements come from (\ref{Q2}).
Figure \ref{fig:alignment_uniformity} depicts the impact of the proposed \scl method on the alignment and uniformity on \tmall and \diginetica datasets.
In general, we find that 
(1) \scl has improved the uniformity of item representations, leading to an improvement in model performance; and 
(2) a higher loss in alignment $\ell_{\text{align}}$ does not necessarily result in worse performance if the uniformity loss $\ell_{\text{uniform}}$ is improved. 

\paragraph{\textbf{Better uniformity of item representations brings substantial improvement in performance.}}
The sub-figure in the centre of Figure \ref{fig:alignment_uniformity} illustrates how \scl improves the uniformity of item representations of \dhcn and \cotrec on \tmall and \diginetica.
This is indicated by a lower uniformity loss when \scl is applied.
The uniformity loss measures the dissimilarity between the item representations themselves and a lower uniformity loss indicates that the item representations are becoming more discriminative and less correlated with each other. 
Specifically, the use of \scl results in a reduction of the uniformity loss of \dhcn from -3.86 to -3.92 on the \tmall dataset, and this improvement is accompanied by an increase in P@10 from 28.65\% to 35.14\%.
%

\begin{figure*}[t!]
  \centering
  \includegraphics[width=\textwidth]{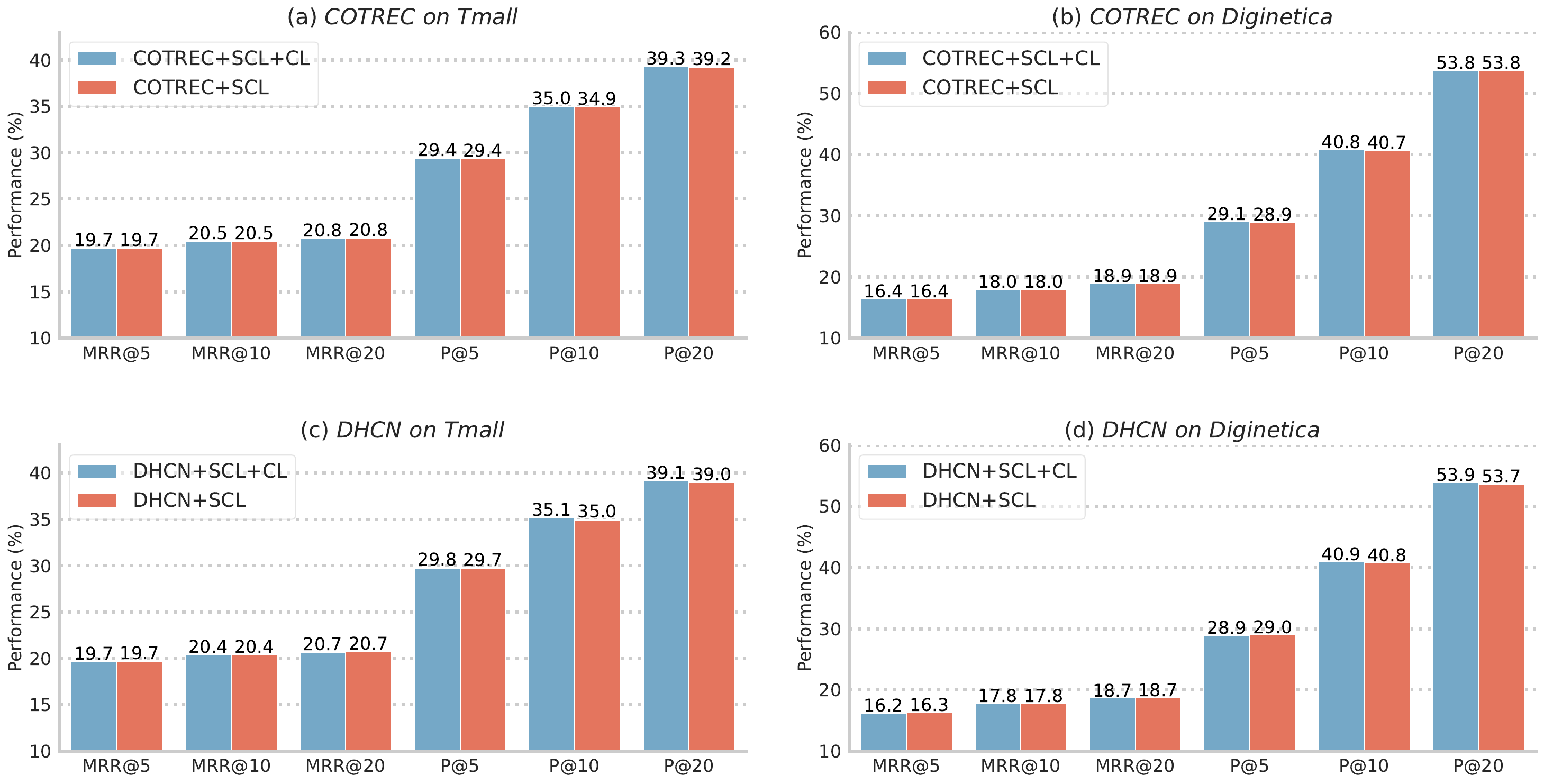}
  \caption{Ablation studies (\ref{Q3}) on other CL objectives.
  Blue indicates the model performance using \scl and other CL objectives. 
  Red represents the model performance using \scl only.
  }
  \label{fig:ablation_contrastive_objective}
  \vspace{-1em}
\end{figure*}

\paragraph{\textbf{The trade-off between alignment and uniformity.}}
We also observe that the proposed \scl method leads to an increase in the alignment loss. 
This indicates that the next item representations are not only becoming more discriminative to other item representations but also less correlated with the session representations. 
We conduct additional studies on the \tmall dataset by adjusting the alignment and uniformity loss through controlling the \scl loss weight $\beta$.
The results are depicted in the two sub-figures of Figure \ref{fig:alignment_uniformity}.
Specifically, for the \dhcn model, when the uniformity loss gradually decreases from -3.86 to -3.92 and the alignment loss increases from 1.08 to around 1.20, the model performance in P@10 is generally improved from 33.62\% to 35.14\%.
Similar results are observed in the experiments using the \cotrec model.
This suggests that an excessive focus on alignment and inadequate attention to uniformity could result in sub-optimal model performance.


\subsection{Sophisticated CL objectives are unnecessary (RQ3)}
\label{sec:really_matter}
Given the complexity of CL objectives used in the SOTA models, we investigate the necessity of these complex contrastive objectives with our proposed \scl approach (\ref{Q3}).
We conduct experiments for \cotrec and \dhcn on two datasets, \tmall and \diginetica with two different settings as follows:
(1) \textbf{\modelscl} refers to the model performance with our \scl and all CL objectives in the original model;
(1) \textbf{\modelsclcl} refers to the model performance with our \scl only.


\paragraph{\textbf{Results.}} 
Figure \ref{fig:ablation_contrastive_objective} depicts the performance of the models. 
It can be observed that \modelscl and \modelsclcl achieve very similar performance results, which implies that the utilization of other sophisticated CL objectives may not be necessary and that the proposed \scl is able to effectively improve the model performance on its own. Below we delve deeper into these findings and discuss them in more detail.
Specially, for the \tmall dataset, 
when using the \cotrec model as the backbone, the \modelscl and the \modelsclcl achieve P@10 scores of 35.0\% and 34.9\%, respectively, and the same M@10 scores of 20.5\%.
Similar performance can be observed for the \dhcn model and \diginetica dataset, implying that the complexity of these objectives may not be necessary when using \scl.

\begin{figure*}[!ht]
  \centering
  \includegraphics[width=\textwidth]{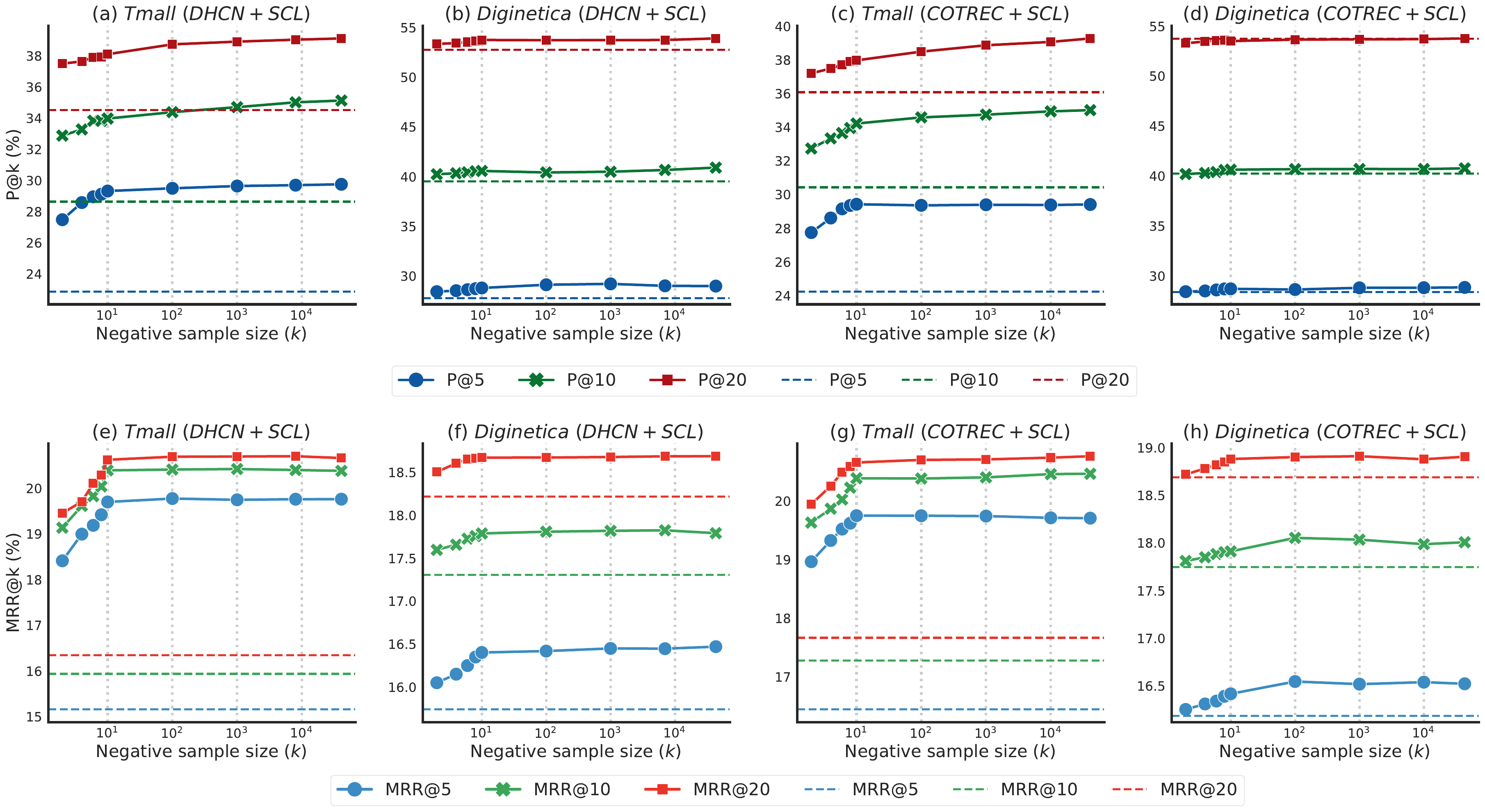}
  \vspace{-1em}
  \caption{The impact of negative sample sizes with \dhcn+~\scl and \cotrec+~\scl on the \tmall and \diginetica datasets (\ref{Q4}).
  The original model performance without using \scl is represented by the corresponding dash line with the same colour. \scl could achieve SOTA performance even when the negative sample size $k$ is equal to 2.
  }
  \label{fig:negative_sample_size}
  \vspace{-1em}
\end{figure*}

\subsection{Computational Cost (RQ4)}
\label{sec:computation_cost}
The size of negative samples plays a critical role in the model performance for CL.
However, it also has some potential drawbacks, including increased computational resources and model complexity, which may make the proposed method impractical for certain applications or settings. 
%
The proposed \scl method has a time complexity of $O(n^2*d)$, where $n$ is the number of item representations used in Eq. \ref{eq:scl} and $d$ is the dimension of item representations. 
To reduce the computational cost, we simplify the objective function of \scl by encompassing a $k$-Nearest Neighbour (kNN) with a fast dense embedding retrieval method, which reduces the time complexity from $O(n^2*d)$ to $O(n*k*d)$, where $k \ll n$.
The updated objective is as follows:
\begin{equation}
    \label{eq:knn}
    \begin{aligned}
        \mathcal{L^{\text{knn}}_{\scl}} = - \sum_{i=1}^n \log \frac{f(\vx_i, \vx_i)}{\sum \limits_{\vx' \in \mathcal{K}_i} f(\vx_i, \vx')},
    \end{aligned}
\end{equation}
where $\mathcal{K}_i$ is a set of $k$ nearest item representations in the distance measured by the cosine similarity for the $i$-th item representation, including its own representation. 
We conduct experiments to evaluate the impact of negative sample size $k$, with the values of 2, 4, 6, 8, 10, 100, $1\,000$, $10\,000$ and the full set.

\begin{figure}[h!]
  \centering
  \includegraphics[width=\columnwidth]{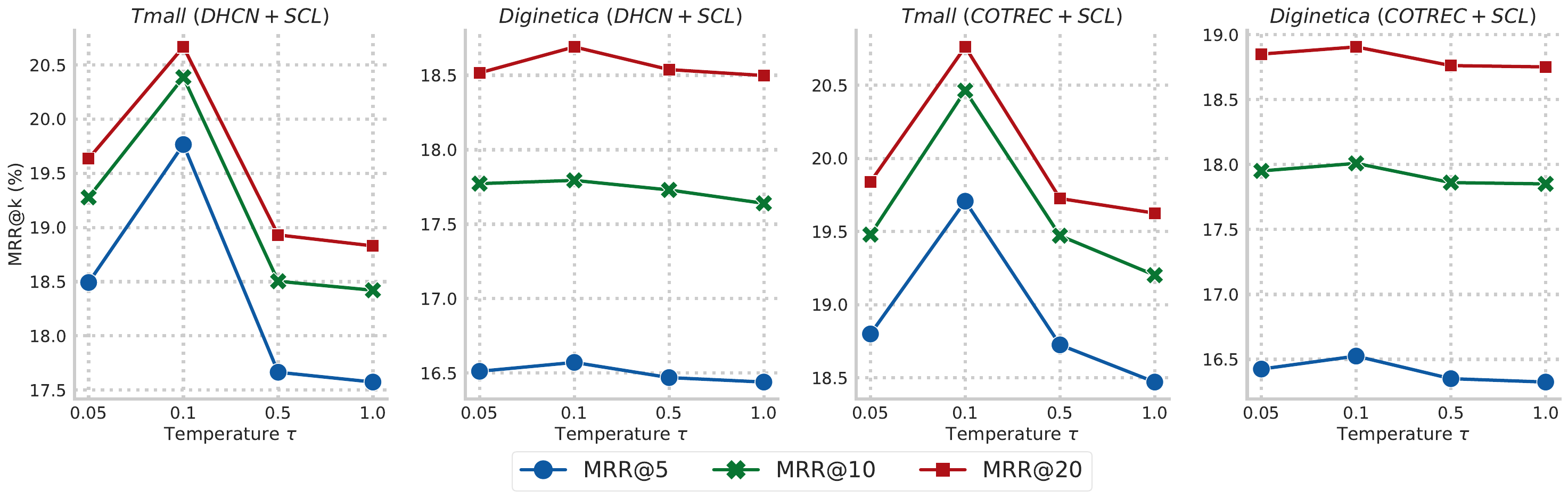}
  \caption{The effect of the temperature $\tau$ using the \dhcn+~\scl and \cotrec+~\scl model on the \tmall and \diginetica datasets, where MRR@$k$ is reported as the representative of model performance.}
  \label{fig:temperature}
  \vspace{-1em}
\end{figure}

\paragraph{\textbf{Results.}}
Figure \ref{fig:negative_sample_size} presents the model performance in P@$k$ and MRR@$k$ with respect to various values of the negative sample size $k$ on the \tmall and \diginetica datasets, where \dhcn+~\scl and \cotrec+~\scl are evaluated.
Overall, our experimental results indicate \scl could improve the performance of SOTA models even when the negative sample size $k$ is equal to 2, and that the performance of the models generally improves as the size of negative samples increases. 
As the negative sample size continues to increase, the improvement of the model tends to level off and become less noticeable, using a small value for $k$ can produce comparable results to using values greater than $10\,000$.
For example, the performance of P@$k$ and MRR@$k$ for the \dhcn+~\scl model tends to become stable once the negative sample size reaches 10 on the \tmall dataset, as shown in sub-figure (a) and (e) of Figure \ref{fig:negative_sample_size}. 
%

%% file: table/baseline.tex
\newcommand\myfonts{\fontsize{8.3pt}{9.96pt}\selectfont}
\newcommand\myfontsize{\fontsize{3pt}{3.6pt}\selectfont}
\newcommand{\up}[1]{\myfonts ($\textcolor{green}{\blacktriangle}#1\%)$}
\newcommand{\down}[1]{\myfonts ($\textcolor{red}{\blacktriangledown}#1\%)$}

\begin{table*}[!t]
\caption{
Performances of all comparison methods on the dev set (\ref{Q1}). 
$\dagger$ from the original paper. 
$\ddagger$ from our own reproductions. 
Triangles in colours indicate an improvement over our reproduced results. 
The highest results in each column are highlighted in bold font.
SOTA performances are indicated in blue.
*** indicates a p-value $<$ 1e-20 for the improvement, ** indicates a p-value $<$ 1e-5, and * indicates a p-value $<$ 1e-2.
}
\label{table:experiment_results_1}
\begin{adjustbox}{max width=\textwidth}
\begin{tabular}{*{13}{c}}
\toprule
\multirow{2}{*}{\bf Method} &
\multicolumn{4}{c}{\bf \large \tmall} & \multicolumn{4}{c}{\bf \large \nowplaying} & \multicolumn{4}{c}{\bf \large \diginetica} \cr
\cmidrule(lr){2-5}
\cmidrule(lr){6-9}
\cmidrule(lr){10-13} 
                                  & P@10   & MRR@10   & P@20   & MRR@20   & P@10 & MRR@10 & P@20   & MRR@20   & P@10   & MRR@10  & P@20        & MRR@20  \\ \midrule
FPMC                              &13.10     & 7.12     & 16.06    & 7.32     & 5.28   & 2.68   & 7.36     & 2.82     & 15.43    & 6.20    & 26.53         & 6.95 \\
GRU4REC                           & 9.47     & 5.78     & 10.93    & 5.89     & 6.74   & 4.40   & 7.92     & 4.48     & 17.93    & 7.33    & 29.45         & 8.33 \\
NARM                              & 19.17    & 10.42    & 23.30    & 10.70    & 13.6   & 6.62   & 18.59    & 6.93     & 35.44    & 15.13   & 49.70         & 16.17 \\
STAMP                             & 22.63    & 13.12    & 26.47    &13.36     & 13.22  & 6.57   & 17.66    & 6.88     & 33.98    & 14.26   & 45.64         & 14.32 \\
SR-GNN                            & 23.41    &13.45     & 27.57    & 13.72    & 14.17  & 7.15   & 18.87    & 7.47     & 36.86    & 15.52   & 50.73         & 17.59 \\
\midrule
$\text{GCE-GNN}^{\dagger}$        & 28.01    & 15.08    & 33.42    & 15.42    & 16.94  & 8.03   & 22.37    & 8.40     & 41.16    &  18.15  & 54.22         & 19.04 \\
$\text{GCE-GNN}^{\ddagger}$       & 27.48    & 14.85    & 32.52    & 15.20    & 17.19  & 8.09   & 22.42    & 8.45     & 40.98    &  18.12  & 54.23         & 19.04 \\
w/ \scl                           & $28.67^{**}$&$15.20^{*}$ & $33.65^{**}$& $15.55^{*}$&\blue$17.44^{*}$&\blue$\bf8.10$&\blue$22.81^{*}$&\blue$\bf8.47$&\blue$\bf41.93^{**}$&\blue$\bf18.45^{*}$&\blue$\bf54.93^{*}$ &\blue$\bf19.38^{*}$ \\ 
$\Delta$ (\%)                     &\up{4.3}  &\up{2.4}  &\up{3.5}  &\up{2.3}  &\up{1.5}&\up{0.1}&\up{1.7}  &\up{0.2}  &\up{2.3}  & \up{1.8}&\up{1.3}       &\up{1.8}   \\
\midrule
$\text{COTREC}^{\dagger}$         & 30.62    & 17.65    & 36.35    & 18.04    & -      & -      & -        & -        & 41.88    & 18.16     & 54.18       & 19.07    \\
$\text{COTREC}^{\ddagger}$        & 30.44& 17.28    & 36.09    & 17.67    & -      & -      & -        & -        & 40.26    & 17.75     & 53.75       & 18.69    \\ 
w/ \scl                           &\blue$35.03^{***}$&\blue$\bf20.46^{***}$&\blue$\bf39.29^{***}$&\blue$\bf20.76^{***}$& -      & -      & -        &  -       & $40.78^{*}$    & $18.00^{*}$ & $53.78$  & $18.90^{*}$    \\
{\myfonts $\Delta$ (\%)}          &\up{15.1} &\up{18.4} &\up{8}  &\up{17.5} & -      & -      & -        & -        &\up{1.3}  &\up{1.4}   &\up{0.1}   & \up{1.1} \\
\midrule
$S^{2}$-$\text{DHCN}^{\dagger}$   & 26.22    & 14.60    & 31.42    & 15.05    & 17.35  & 7.87   & 23.50    & 8.18     & 39.87    & 17.53   & 53.18         & 18.44 \\ 
$S^{2}$-$\text{DHCN}^{\ddagger}$  & 28.65    & 15.94    & 34.54    & 16.35    & 17.23  & 7.70   & 23.00    & 8.10     & 39.54    & 17.31   & 52.76         & 18.22 \\
w/ \scl                           &\blue$\bf35.14^{***}$&\blue$20.39^{***}$&\blue$39.13^{***}$&\blue$20.67^{***}$&\blue$\bf17.61^{*}$&\blue$7.92^{*}$&\blue$\bf23.74^{**}$&\blue$8.32^{*}$ & $40.91^{**}$ & $17.79^{**}$ & $53.91^{*}$ & $18.69^{*}$ \\
{\myfonts $\Delta$ (\%)}          &\up{22.7} &\up{27.9} &\up{13.3} &\up{26.4} &\up{2.2}&\up{2.9}&\up{3.2}  &\up{1.7}  &\up{3.5}  &\up{2.8} &\up{2.2}       &\up{2.6} \\
\bottomrule
\end{tabular}
\end{adjustbox}
\end{table*}

%% file: paper/6_experiment_rq4.tex
\subsection{Hyperparameter Sensitivity}
We conduct an additional study to investigate the effect of varying the hyperparameter temperature $\tau$ on model performance. 
In the experiment, 4 distinct values of $\tau$ (namely 0.05, 0.1, 0.5, and 1.0) are evaluated with the \dhcn+~\scl and \cotrec+~\scl on the \tmall and \diginetica datasets.
The experimental results are presented in Figure \ref{fig:temperature}, indicating that the model achieves optimal performance when the temperature $\tau$ is set to 0.1.

%% file: paper/8_conclusion.tex
\section{Conclusion}


In this work, we propose \textit{Self Contrastive Learning} (\scl), which improves the performance of SOTA models with statistical significance across three datasets.
\scl targets the optimization of item representation uniformity in SOTA session-based recommendation systems. 
\scl complements the use of cross-entropy loss, eliminating the need for sophisticated CL objectives.
This simplicity makes \scl highly adaptable across a variety of models.
Moreover, we shed light on how \scl enhances representation spaces from the alignment and uniformity viewpoints, thus emphasizing the importance of uniformity in item representations.
Our analysis also points out that achieving an optimal balance between alignment and uniformity loss is a crucial aspect of designing recommendation systems.
Lastly, we demonstrate that the implementation of \scl is efficient and entails low computational costs.